# *Using the medical image processing package, ImageJ, for astronomy*

By Jennifer L. West and Ian D. Cameron

*Abstract:* At the most fundamental level, all digital images are just large arrays of numbers that can easily be manipulated by computer software. Specialized digital imaging software packages often use routines common to many different applications and fields of study. The freely available, platform independent, image-processing package ImageJ has many such functions. We highlight ImageJ's capabilities by presenting methods of processing sequences of images to produce a star trail image and a single high quality planetary image.

*Introduction*

ImageJ is a particularly flexible image-processing package that is a free, open source program developed by Wayne Rasband at the National Institutes for Health (NIH). The program will run on all computer platforms including Windows, Mac, Linux, Unix, and even the Sharp Zaurus PDA. ImageJ claims to be the fastest pure Java image processing program; it is capable of filtering a 2048x2048 image in 0.1 seconds (according to the ImageJ website).

ImageJ's intended purpose is medical image processing, but it works very well in an astronomical context. The software has a particular emphasis on analysis and incorporates a number of tools for measuring images. As a result functions are often more mathematically "transparent" than in commercial packages such as Adobe Photoshop.

A well-developed "plugin" interface allows for customization. A very active user community has contributed dozens of freely available plugins, some of which are very useful for astronomy (see appendix). There are even plugins available to interface with various CCD cameras and webcams. J. West has produced a plugin to interface with an SBIG CCD camera and various others are available on the ImageJ website with new ones frequently being added. A "macro" interface allows tasks to be recorded and automated.

This paper will highlight the functions in ImageJ that are useful for astronomy both in terms of measurement of astronomical data and for the creation of "pretty pictures". Two "case studies" of applications of ImageJ's usefulness in an astronomical context are presented. Case study 1 is a straightforward application of ImageJ's capabilities. We demonstrate how ImageJ can produce a time-lapse movie and a star trail image from a set of digital photos. Case study 2 is a more complex application. Here we undertake an experimental approach to process a webcam movie of a planet to produce a single, high quality image.

*ImageJ Commands*

*Image input and output –*

ImageJ can read and write a wide variety of image file formats including TIFF (uncompressed), GIF, JPEG, BMP, PGM and FITS (writing FITS files is accomplished through a plugin contributed by J. West). The program can also read and write raw data (the user must know the dimensions and image data type) and text images. There are currently no publicly available plugins to open raw files produced by commercial digital cameras. Of particular interest to astronomy is the ability to load sequences of images to create what ImageJ calls a "Stack." Stacks are discussed in further detail later in this paper.

In addition to single frame images and image sequences, ImageJ can also load movies in AVI or QuickTime formats. User contributed plugins are also available to read and write other formats such as animated GIFs, PNGs and PICTs among others.

*File types –*

ImageJ is capable of handling both grayscale and colour images. Permissible file types include 8-bit, 16-bit, and 32-bit grayscale as well as 8-bit colour, RGB colour and HSB colour. Other colour spaces such as LAB are supported through user-contributed plugins.

*Working with Image Stacks –*

Similar in concept to "layers" in Adobe Photoshop or Gimp, ImageJ uses "stacks" to work with a group of images at one time. The individual images are called "slices". Available computer memory is the only limit to the number of slices that can be included in the stack.

Stacks are very versatile and useful for several different applications. Movies in AVI, QuickTime, or animated GIF formats can be loaded as a stack. In addition, a directory of images can automatically be opened as a stack. Stacks can also be read and written as multi-image TIFF or RAW files. Thus, 3D images such as multi-channel radio data can be visualized.

Standard image manipulation (such as level adjustments and filters) and measurements can be performed on the images individually, or on the stack as a whole. This greatly simplifies the task of batch processing.

The images in a stack can be viewed in several ways. Animation controls allow the images to be displayed as a movie. The data can be projected or plotted along any axis and the user is given the option of the "projection type", which includes such options as "Average", "Maximum", and "Median". The 3D "cube" can be rotated and animated as well.

There is no built in functionality for registering images in ImageJ. There is however, an excellent set of plugins called "TurboReg" and "StackReg", contributed by Thévenaz, Ruttimann & Unser (1998) to perform this task.

StackReg accepts a stack of images as input (such as a movie that has been loaded into the program) and automatically aligns each image to the selected slice. The user is given the option of five standard types of registration algorithms: Translation, Rigid Body, Scaled Rotation, Affine, and Bilinear.

*Image Selections –*

The ImageJ toolbar (**Figure 1**) provides tools for selecting regular and irregular areas (called Regions Of Interest or ROIs) on an image. Several types of selections such as rectangles, circles, poly-line, and a "magic wand" are available. Selections can be measured, filtered, filled or drawn.

A tool called the "ROI Manager…" (located under the "Analyze" -> "Tools" submenu) may be used to store, combine, and save multiple selections at one time. The ROI Manager may be used to measure the same region on multiple images or on all of the images in a stack.

*Colour and level adjustment –*

Basic brightness/contrast and minimum/maximum level adjustments are available in ImageJ. This includes the allowance for colour adjustments by separately manipulating R, G, and B (as well as C, M, and Y).

Several lookup tables (or "LUTs") are also included in the software and allow for customization over how the data is displayed without changing the actual pixel values in the image. Hundreds of additional LUTs may be downloaded from the website. A plugin, contributed by J. West, allows a LUT to be defined using standard mathematical functions such as Log and Square Root.

*Other image adjustments –*

A range of standard image adjustments such as rotation, resizing (with or without interpolation), cropping, duplicating, zooming, and renaming are fully supported.

*Image Filtering –*

Several standard image filters are included in ImageJ. Some of these include "Gaussian Blur", "Median", "Mean", and "Unsharp Mask", among others. There is also the built in capability of specifying a user-defined convolution mask. Dozens of user-contributed plugins provide access to many other filters including Wavelet filters. Fourier transforms and frequency filtering are also part of the standard ImageJ package.

A wide variety of mathematical functions may be easily performed on image pixels. For example, one may add, subtract, multiply, or divide by a constant, perform logarithms, square root, reciprocal, or find the absolute value of a selection of pixels in an image.

*Image Calculator –*

The "Image Calculator" allows images to be combined using one of several mathematical functions, including add, subtract, multiply, and divide as well as AND, OR, XOR, min, max, and average. This is useful for performing calibrations such as dark subtraction and flat fielding on astronomical images. The image calculator accepts stacks as input, thus a stack of image can be batch reduced.

*Measuring Images –*

A wide variety of measurements including Center of Mass, Min/Max Gray Value, Mean Gray Value, Standard Deviation, and Area can be performed on an image or a stack of images.

The measurements that should be performed are specified by the "Set Measurements…" command, located under the "Analyze" menu. The "Measure" command (also under the "Analyze" menu) can then be executed to measure a single image. In order to measure the entire stack at once, the "ROI Manager…" must be used (see Section on *Image Selections*).

Measurements appear as a table in a separate results window. The measurements can be copied and pasted directly into a spreadsheet where they can be sorted.

*Image Display Tools –*

Image scale, for example arcminutes per pixel, can easily be set using ImageJ's "Set Scale…" command. A scale and/or grayscale calibration bar can also easily be drawn on the image as shown in **Figure 2**.

*Macros and Plugins –*

More than 300 plugins are freely available on the ImageJ website. Plugins are simple to install. All that is required is to download the file(s) to ImageJ's "plugins" directory and then restart the program. Plugins that are placed in a subfolder within the plugins folder will show up on a submenu when the program is launched. Plugins are usually written in the Java programming language, but other languages such as JPython are also supported. The interface is relatively straightforward for someone familiar with Java programming. A useful "Compile and Run" feature is included within the program, which greatly speeds up development, as there is no need to continually restart the application while debugging code. An online plugin writing tutorial is also available.

For those not comfortable with programming, a macro scripting language allows a series of commands to be automated. A "Record" function simplifies the task of writing a macro. More than 200 macro examples can be downloaded from the ImageJ website.

***Case study 1 – Creating a Star Trail Image from a Sequence of Short Exposures:***

With today's digital cameras, it is difficult to capture star trails since exposure times longer than a few minutes will cause the background to wash out the image. The solution is to take a sequence of short exposures and combine them together. ImageJ allows this to be done very easily.

The sequence of images should first be opened as a stack. Put all of the files in a single directory and ensure that the sequential order of the files names corresponds to the order that you wish to open the images. Next, under the "File" menu, select "Import" -> "Image Sequence…" Select the first image in the directory and click "Open". A dialog box with several options should be displayed. This dialog box allows the user to select the range of images that should be opened. There is also the option to increment the sequence, for example to open every second image in the directory. The images may also be scaled and/or converted to grayscale or RGB. Click "OK" to import the sequence.

The next step is to adjust the brightness and contrast or apply any filters as may be required. The same settings can be applied to all of the images or just a single frame. Individual frames can be deleted from the stack by selecting "Delete Slice" under the "Image" -> "Stacks" submenu.

The final step is to combine all of the frames. We will use the "Z Project…" command, which is also located under the "Image" -> "Stacks" submenu. To create a star trail image keeping the background as dark as in a single frame, we will use the "Max Intensity" projection type. This projection method will create a new image by selecting the pixel with the maximum intensity for each x-y grid point in the stack.

The disadvantage of this method is that noise will be accentuated. If the original frames are excessively noisy, then steps should be taken to reduce noise before combining the frames. One possible method would be to average a few frames at a time and then combine the averaged frames. However, if the original images are well exposed then this method can produce a very striking star trail image.

The sequence of images in the stack can also easily be saved as a time lapse movie in either AVI or QuickTime formats. These formats are both available under the "Save As…" menu. In addition, a user-contributed plugin is available to save the stack as an animated GIF.

**Figure 3** shows the final combined image produced by using this method on a sequence of 100 40-second exposures with a 30-second gap between each exposure. The camera used was a Canon 20Da with a 15 mm lens at f2.8 and an ISO speed of 1600. The images were acquired over a period of approximately 2-hours, between 12:06 AM and 2:01 AM CDT, in the early morning of Saturday, September 2, 2006 by Jennifer West. The location was Spruce Woods Provincial Park in South Western Manitoba during the Spruce Woods Star Party.

*Case study 2 – Processing a Sequence of Planetary Images:*

Using a webcam to produce high-quality planetary images is a well-known technique in the world of amateur astronomy. A popular freeware program called "Registax" allows the user to combine and process individual frames from a webcam movie. Though this program does an excellent job there are some disadvantages. Firstly, the program is only available for Windows; and secondly, the program's interface is somewhat restrictive. With ImageJ we can more easily customize and extend our processing procedure in response to the available data.

With a thoughtful selection of algorithms and parameters, we are able to use ImageJ to produce an image as good, if not better than an image created with the same data using Registax. This case study discusses our method in detail and we include rules-of-thumb to allow the reader to make intelligent parameter choices when creating ones own image.

We processed a sequence of 451 images of Jupiter taken with a Philips ToUcam Webcam on the night of April 9, 2004 through a Celestron 11-inch telescope. The data were kindly provided by Jay Anderson.

The general stages of processing a sequence of planetary images includes loading the raw data, registering (aligning) the individual frames, normalizing the stack, selecting the "best" frames, and enhancing and combining the images.

*Loading raw data –*
Many webcams output movies in AVI format, which can be directly loaded into ImageJ as a stack of single frame images. The user can view all "slices" in the stack by using a scroll bar or by "animating" the stack.

*Registering (aligning) the individual frames –*
Using the plugin "StackReg", the images can be automatically registered. Using the Rigid Body rotation option, which allows translations and rotations but maintains the size of the original image, works very well for a sequence of planetary images that are obtained with a webcam.

Once the frames have been aligned, crop the stack to the desired size. The easiest way to do this is to select a square or rectangular region using the rectangle selection tool then choose "Crop" from the "Image" menu.

*Normalizing the stack –*
If there are any differences in the brightness values of the slices in the stack due to effects such a changing sky conditions these must be normalized before moving on to selecting the best frames. A plugin, called "Normalize Values" was written by J. West, to achieve this task. The plugin measures the mean value of each frame and finds the highest mean value in the stack. Each image in the stack is then multiplied by a value that is equal to the highest mean/mean of the slice. This raises the mean value of each image to a value equivalent to the highest mean.

*Selecting the "best" frames –*
One of the more challenging aspects of processing a stack of planetary images is selecting the best frames. This is especially true when faced with selecting from hundreds of images.

Since the salient features of Jupiter are the cloud belts, we use an edge detection algorithm to select the images that have the most well defined edges. This algorithm is built into ImageJ as the "Find Edges" filter, located under the "Process" menu. This command convolves the image with two standardized Sobel filters, which calculate the derivatives of the image in both vertical and horizontal directions (see Gonzalez & Woods, 2002 for further explanation). A single filtered image is produced that shows the edges as bright regions and the remainder of the image as dark regions, as shown in **Figure 4**. The mean value of the entire filtered image is then calculated. Planet images that have sharper, more discontinuous edges will produce a filtered image with a higher mean value than images that have blurry or soft edges.

The "Find Edges" command may be executed on the stack as a whole, producing a stack of filtered images (the original images will be over-written, so it may be helpful to first execute the "Duplicate…" command located under the "Image" menu). The mean value of each image in the stack can then be measured.

In addition to measuring the stack of filtered images, one can also sort it by the mean value of the slices. This is accomplished by using a very useful plugin called the "Stack Sorter", written by Bob Dougherty.

After the stack has been measured and/or sorted, the decision as to how many frames to use for the final image must be made. Enough images must be used to ensure good signal to noise. However, using too many images will smooth out the final image and fine detail may be lost. This is especially true if poor quality images are included in the final selection. In general, use as few images as possible to retain a sufficient amount of signal to noise. This effect is illustrated in **Figure 5**. The image created by combining eight images, although noisier, contains a significantly greater amount of detail than the image created by combining 451 images. A compromise between these two extremes would probably be the ideal solution. For the final image (**Figure 6**), we used the best 28 images and found this to be the most pleasing image visually.

A plugin called "Select Frames With Best Edges" was written by J. West to automate this whole procedure. This plugin takes a user input of a threshold value. It then creates a new stack of the best images by comparing each image to the best image and includes those that are no worse than the threshold value.

*Enhancing the Images –*

Although the photometry is not preserved, unsharp masking is a very effective technique for extracting detail in the particular context of processing planetary images. The algorithm of unsharp masking involves subtracting a blurred version of the image from the original image. In this way, some of the larger scale (lower frequency) component of the image is removed, thereby enhancing the smaller scale (higher frequency) component. A severely unsharp-masked image will appear artificial and over-processed and may reveal "structure" that is in fact an artifact created by the processing.

We perform an unsharp mask filter on the individual raw images before combining the frames into the final image. We find this technique allows a more severe unsharp mask filter to be used initially. By combining the images afterwards, the effects of the filter are averaged and the final-result looks more natural.

ImageJ's built in "Unsharp Mask" filter blurs the image by convolving with a Gaussian function with a user specified radius. The blurred image is then scaled by a user-specified factor before it is subtracted from the original. Selecting the appropriate factors is a combination of science and art.

In an image of a planet, the detail in which we are usually most interested in high spatial frequencies, i.e. we are interested in the rapidly varying structure on the disk. Unfortunately, this is also the scale where we find noise. By using unsharp masking, we can enhance the high frequencies where the interesting structure is located but we also enhance noise. The trick is to combine multiple images by averaging, which will reduce the noise that varies from frame to frame, but retain the high-frequency detail that should be present in each frame. The rule-of-thumb that we use is to select the radius of the Gaussian filter such that it is a little larger than the scale of the features that one is attempting to enhance. For example, for an image of Jupiter, we would choose a blurring radius that is approximately equal to the width of one of the cloud bands. By subtracting off an image blurred at this scale, we are reducing the contribution of all features larger than that radius and enhancing the contribution of all features that are smaller.

*Combining the Images –*

Before the images are combined, it may be desirable to run StackReg once more and refine the image registration. The last step is to combine the images in the stack using the "Z Project…" command. The most useful projection type for the purposes of combining planetary images is "Average."

*Conclusions –*

ImageJ is a very versatile base for a wide range of image processing applications. Our goal in this paper was to demonstrate the large range of current capabilities and to make the reader aware of the great future potential that this software offers. It is our hope that it will be embraced by a growing number of professional and amateur astronomers as a viable alternative to other processing and visualization packages.

*References –*

## *Appendix*

### *Useful Plugins –*

Following is a listing of some of the plugins that we find useful for analyzing or processing astronomical images. Links to each of the individual plugins as well as a downloadable zip archive of the complete set can be found at:
http://www.umanitoba.ca/faculties/science/astronomy/jwest/imagej.html

### *TurboReg and StackReg, Philippe Thévenaz, Biomedical Imaging Group, Swiss Federal Institute of Technology Lausanne*
TurboReg registers, or aligns, two or more images. The companion plugin, StackReg will register all of the image slices in a stack to the top image. Alignments can be performed automatically or manually by selecting landmarks in both the source and target images. The user is given the option of five alignment methods: translation, rigid body, scaled rotation, affine, and bilinear.

### *MosaicJ, Philippe Thévenaz, Biomedical Imaging Group, Swiss Federal Institute of Technology Lausanne*
This plugin is used to create composite mosaic images, such as panoramic images, from any number of overlapping images. The input images do not have to be in a grid pattern; any regular or irregular arrangement of images may be used as long as there is sufficient overlap. The TurboReg engine (see above) is used for image registration and therefore, it must also be installed.

### *Contour Plots, Walter O'Dell, University of Rochester*
Draws contours at levels specified by the user. The contours are non-destructive but they may also be drawn on the image if desired. Contour lines can be converted to selections and the area within can be measured using ImageJ's standard measurement tools.

### *Interactive 3D Surface Plots, Kai Uwe Barthel, Internationale Medieninformatik*
Creates 3D (x, y, and intensity) surface plots that can be scaled and rotated. Plots can be drawn using dots, lines, mesh, or filled. The colours and lighting of the plots may also by dynamically adjusted.

### *Animated Gif Writer, Ryan Raz*
Saves a stack as an animated GIF.

### *Animated Gif Reader, Kevin Weiner (FM Software), Wayne Rasband*
Opens an animated GIF file as a stack.

### *Stack Combiner, Wayne Rasband*
Creates a new stack by combining two arbitrarily sized input stacks.

*Substack Maker, Anthony Padua, Daniel Barboriak, Duke University Medical Center*
Extracts selected slices from a stack to create a new stack.

*Stack Sorter, Bob Dougherty*
This plugin allows the slices in a stack to be sorted by the mean value of the slices or by label name.

*SaveAs FITS, Jennifer West*
Saves an image in FITS format.

*Plot FWHM, Jennifer West*
Fits a Gaussian function to a horizontal and vertical cut that is centred on the brightest point in the image. Returns the FWHM of the function both in numerical and graphical formats.

*Background Extractor, Jennifer West*
Extracts the background of a stack of images by normalizing each image in a stack and then combining the stack. Used to produce sky flat images from a sequence of exposures.

*Moon Crater Calculator, Jennifer West*
Calculates the height of a crater wall or lunar mountain using user input from measurements on lunar images.

*Select Frames With Best Edges, Jennifer West*
Selects the frames with the "best" edges from a stack of images. The edges are computed using the "Find Edges" routine in ImageJ (using a Sobel filter). The mean values of the edge frames are computed and the frames are ordered according to this value. A new stack is created that contains all the frames having a mean value that is at least xx% as high as the highest mean value.

*Normalize Values, Jennifer West*
Measures the mean value of all images in a stack and finds the highest mean. Each image is then multiplied by a constant equal to the highest mean/mean of the image. This results in each image of the stack having the same mean value.

*Remove Hot or Cool Pixels, Jennifer West*
Replaces all pixels above and/or below a threshold value with either the nearest neighbour median, nearest neighbour mean, or zero.

*Apply LUT from Function, Jennifer West*
Uses a user-defined transformation function to create a new non-linear LUT and apply it to the image. The image display is automatically updated however the data values remain unchanged. Included functions are A $log(Bx)$, A $sqrt(Bx)$, A $exp(Bx)$, A $x^B$, and A $sin(Bx)$, where A and B are user defined constants.

## *Figures*

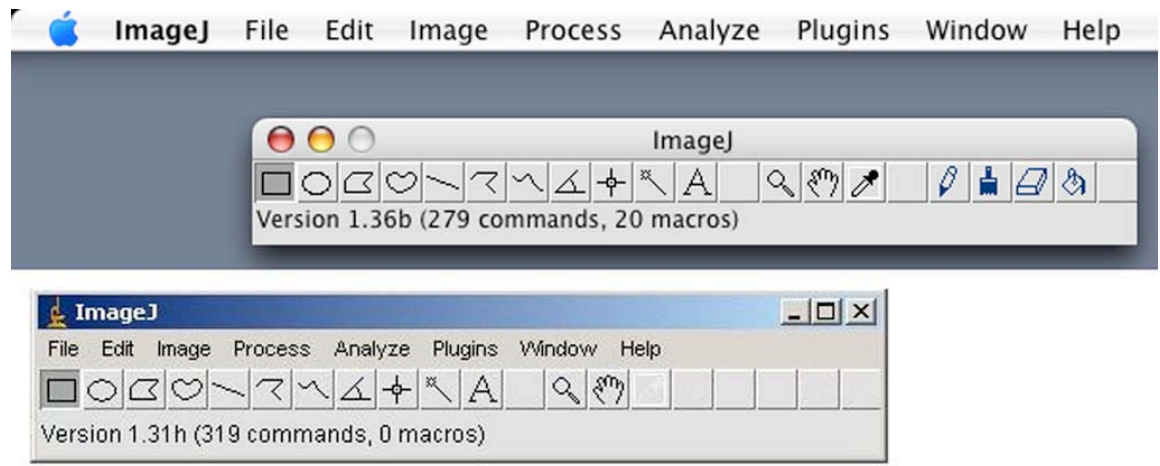

Figure 1: ImageJ toolbar as it appears on both Mac OSX (top) and Windows (bottom) operating systems.

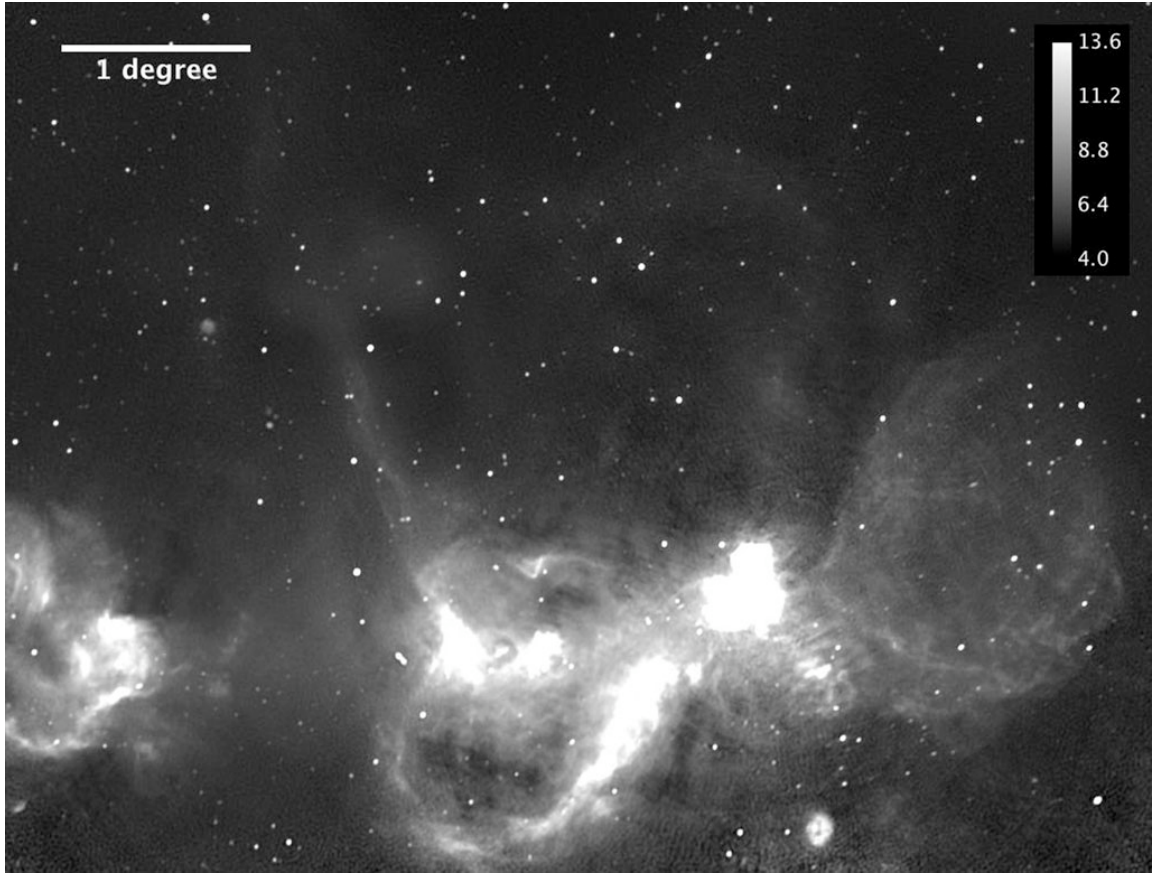

Figure 2: Scale/grayscale calibration bar displayed on a 1420Mhz radio image of the W4 star-forming region (also known as the "Heart Nebula" at optical wavelengths) from the Canadian Galactic Plane Survey (Taylor et al. 2003). The greyscale displayed is Brightness Temperature in units of Kelvin.

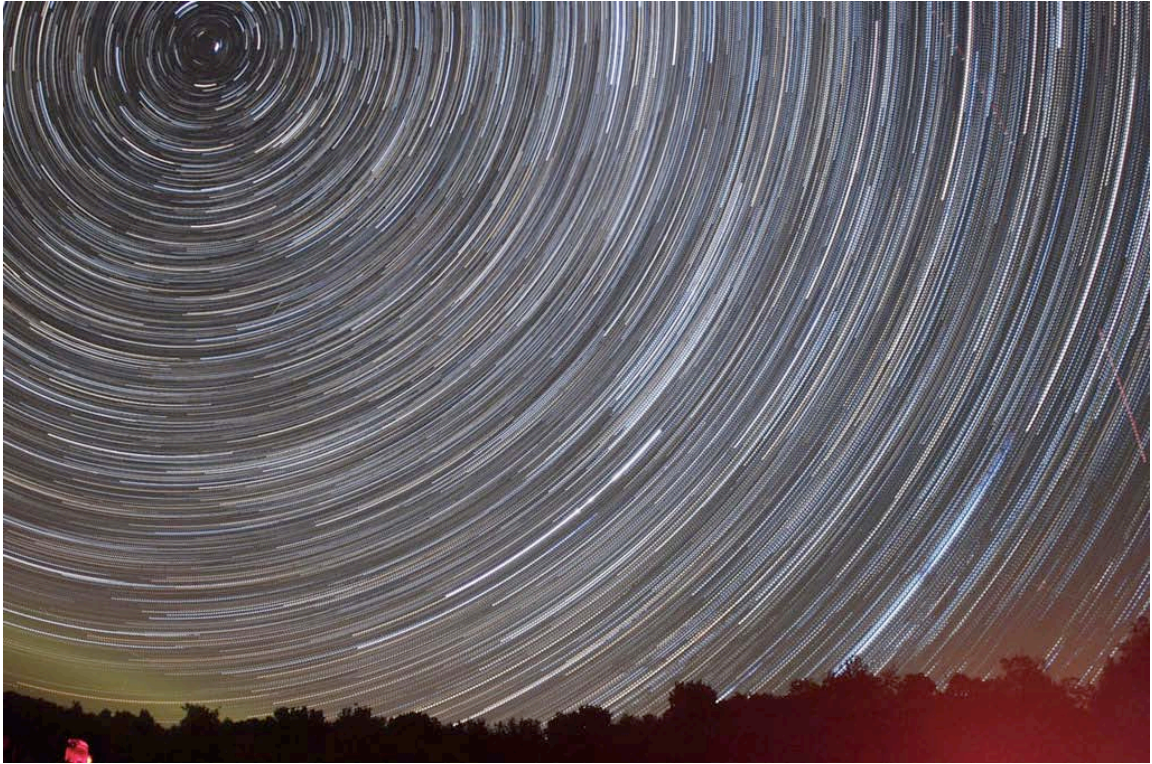

Figure 3: Final star trail image produced from 100 40-second exposures with a 30-second gap between each exposure taken with a Canon 20Da with a 15 mm lens at f2.8 and an ISO speed of 1600.

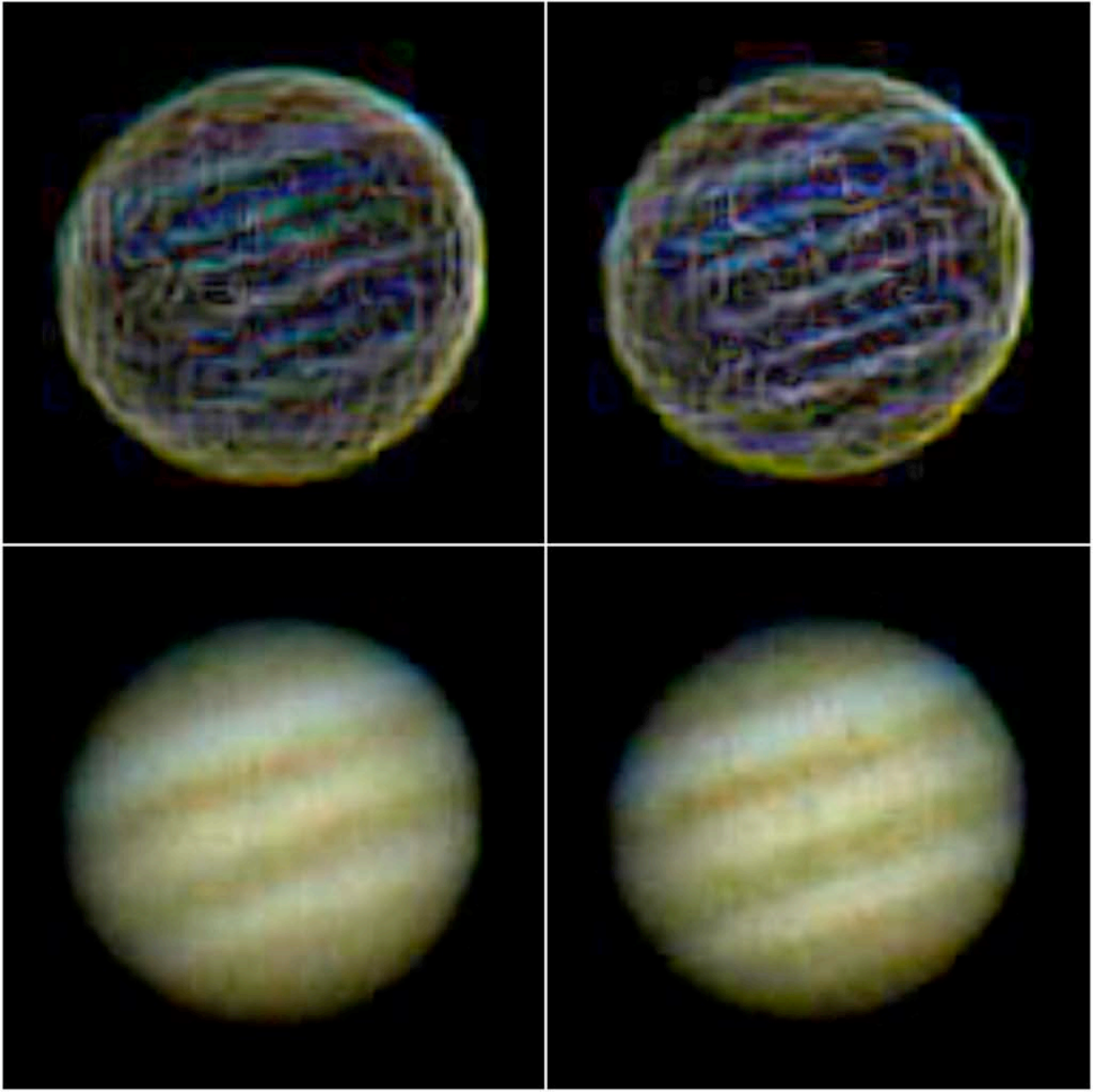

Figure 4: Best (right) and worst (left) individual frames from the Jupiter webcam sequence. The edge-detected images (top) are shown with the corresponding original images (bottom). The images on the right have more clearly defined and brighter edges, especially along the cloud bands.

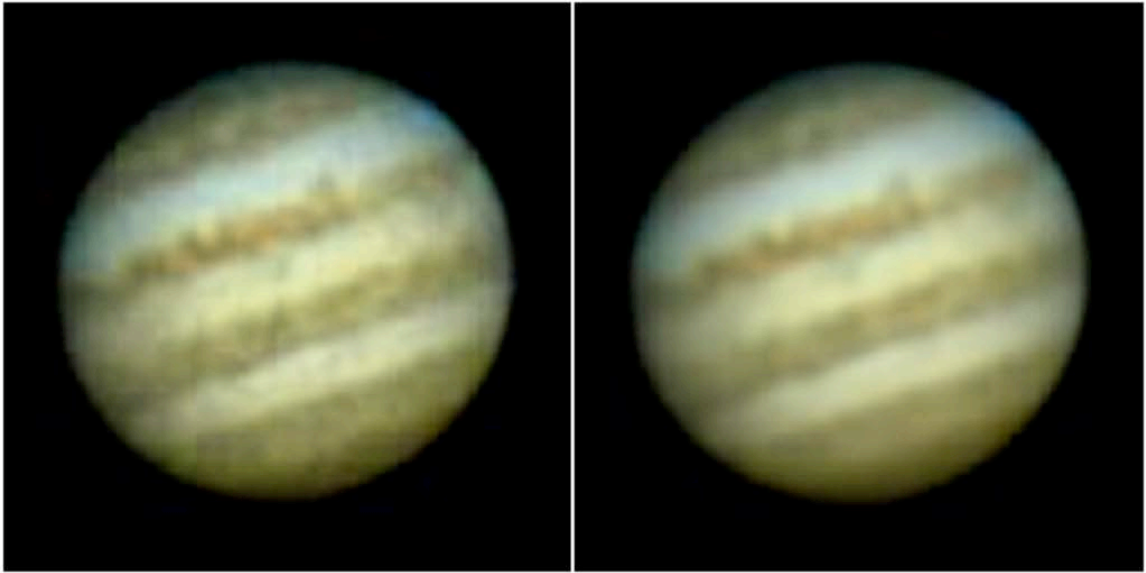

Figure 5: Results of combining the top eight frames (left) and 451 frames (right). The image on the left, while noisier, shows a much greater amount of detail than the image on the right, which has been smoothed because of combining a large number of images.

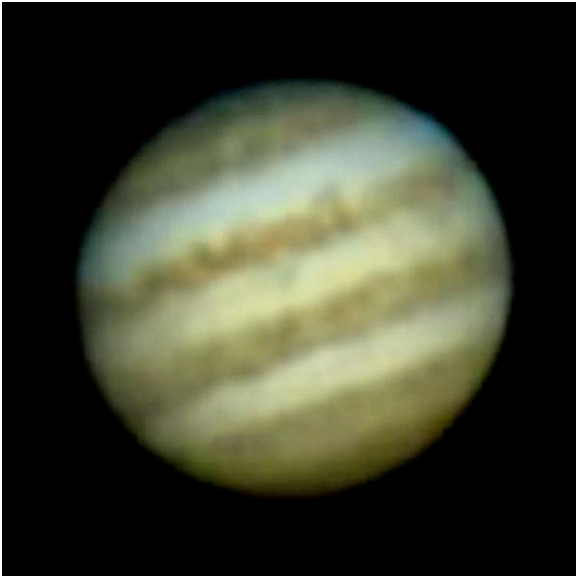

Figure 6: Final image of Jupiter produced by combining the top 28 frames from the webcam movie.